\documentclass[prb,twocolumn,showpacs,preprintnumbers,amsmath,amssymb,superscriptaddress]{revtex4}
\usepackage{graphicx}
\usepackage{dcolumn}
\usepackage{bm}

\begin{document}

\title{Disorder effects in the AHE induced by Berry curvature}

\author{N.A. Sinitsyn}
\affiliation{Department of Physics, University of Texas at Austin,
Austin TX 78712-1081, USA}
\author{Qian Niu}
\affiliation{Department of Physics, University of Texas at Austin,
Austin TX 78712-1081, USA}
\author{Jairo Sinova}
\affiliation{Department of Physics, Texas A\&M University, College
Station, TX 77843-4242, USA}
\author{Kentaro Nomura}
\affiliation{Department of Physics, University of Texas at Austin,
Austin TX 78712-1081, USA}

\begin{abstract}
We describe the charge transport in ferromagnets with spin orbit coupled Bloch bands by combining the 
wave-packet evolution equations with the classical Boltzmann equation.  
This approach can be justified in the limit of a smooth disorder potential. 
Besides the skew scattering contribution, we demonstrate how other effects of disorder appear
 which are closely linked to the Berry curvature of the Bloch states associated with the wavepacket. 
We show that, although being of the same order of magnitude as the clean 
limit contribution, generally
disorder corrections depend differently on various parameters and can lead to the sign reversal 
of the Hall current as the function of the  chemical potential 
in systems with a non-constant Berry curvature in momentum space. Earlier conclusions on the effects
of disorder on the anomalous Hall effect depended stricly on the lack of momentum dependence of 
the Berry curvature in the models studied and generalizations of their findings to other systems with more complicated
band structures were unjustified.
\end{abstract}

\maketitle

\section{Introduction}
The theory of the anomalous Hall Effect (AHE) has a long history
The appearance of the Hall current requires breaking of some basic symmetries. 
It was proposed by Karplus and Luttinger in early 50s \cite{Lutt} that the anomalous Hall effect in ferromagnets results from the interplay
 of the exchange field, which breaks the time-reversal symmetry, and the
spin-orbit coupling, that violates the chiral symmetry. Interestingly, at the same time a similar effect was 
predicted and explained in geometrical optics \cite{Fedorov}, however, its relation to the 
AHE was revealed only recently \cite{{Onoda},{Bliokh}} after the modern interpretation of both effects in terms of the Berry phase, which affects the motion of wave packets, had been constructed.\cite{Jungwirth,Niu,Zeldovich} Luttinger \cite{Luttinger} built a detailed theory
of the AHE based on the high order quantum Boltzmann equation calculations.\cite{Kohn}  In that work Luttinger identified various contributions, known today
as Berry phase contribution, the skew scattering, the side jump on the impurity potential and a contribution that involves interference from many scatterers. 

Since then, a number of theoretical works appeared that extended the theory. However untill recent time 
most  of them had been devoted to what is today called the extrinsic AHE. In the extrinsic AHE a simple Bloch band structure of the system is assumed and the spin orbit interaction is
localized on the impurity potential via terms of the form $\lambda_{SO}\hat{\sigma}_z  {\bf \hat{z}}{\bf k \times \nabla}V_{dis}(x)$ e.t.c. It was recognized by
 Smit, Berger and others 
\cite{{Smit},{Berger1},{Berger2},{Berger3},{Bruno-Dirac},{Extrinsic1},{Extrinsic2}} that in this case the main contributions to
 the Hall current will be those from impurities via the side jump and the skew scattering mechanisms. 
In contrast to the extrinsic AHE, the intrinsic one assumes that the spin orbit coupling is already present in the band structure of the system
and generally cannot be considered as weak in comparison even with the Fermi energy. Recently the interest toward the intrinsic AHE has grown up
considerably due to new applications in the diluted magnetic semiconductors (DMS) and due to the interesting modern interpretation of the Hall current
in terms of the Berry phase.\cite{{Jungwirth},{Fang}} The Berry phase contribution to the Hall current was shown 
to be in good quantitative agreement with experiment in many different materials with strong 
spin-orbit couping in their band structure, giving weight to the theory involving this contribution alone. \cite{{Fang},{Experiment1},{numerics1}}

The Luttinger's theory of the intrinsic AHE, however, predicts that other contributions should arise due to the scattering from
 a disordered potential even  if the disorder potential itself is spin independent. The asymmetry in scatterings  becomes transparent 
 in the basis related to the Bloch states.
As an example, Luttinger considered a rather simplified model which demonstrated that such corrections must reverse the sign of the Hall current in the dc-limit, in comparison to the clean limit, though in the high frequency AC-case the clean limit contribution should dominate.
The same results have been shown by several other works utilizing different approaches and focusing on this simplified model.\cite{{Adams},{Nozieres},{Chazalviel}}  

The results from these simplified models seem to be in a contradiction with recent work which did not find such a change of sign in
numerical simulations or in a comparison with the experimental results.  We show here that a possible resolution of this discrepancy is the simplicity of the early models from which many generalizations were stated without justification and whose results depended drastically on the simple momentum depencence of the Berry curvature of the Bloch states. 

The work by Luttinger and works of other authors, related to the disorder contribution to the AHE, 
are rather involved. Generally various contributions separately
turned out to be not gauge invariant and only the final result was physically meaningful.
Because of these shortcomings, we reformulate the  basic arguments of the previous authors in terms of wave packet dyanamic equations which are
fully gauge invariant and consistent with prior results.\cite{Niu} 
Being gauge invariant, the wave packet equations allow one to identify the physical meaning of various contributions. We should note, however, that the wave packet equations are valid only in the limit of smooth potentials. This restricts the applicability of our conclusions and generalizations
of our results should not be made to regimes where the semi-classical treatment is not justified
without careful checks; therefore we did not make the goal to construct the final theory of 
the disorder in the AHE but rather to construct a simple formalism that demonstrates basic features
of the problem and highlights a basic key ingredient missed by prior theories studying the effects
of disorder, namely the importance of the momentum dependence of the Berry curvature.

We have found that the change of the sign of the Hall current due to the disorder, 
found in the early works, is not universal and arises in the models where  bands have a
constant Berry curvature in the momentum space. 
In diluted magnetic semiconductors the Berry curvature is strongly momentum dependent. 
This separates parametrically the clean contribution from the others and makes the 
sign reversal not universal. Rather generally we find the same sign of the total Hall current
for realistic bands except in extreme situations where one of the bands becomes depleted.

We organize the rest of the paper as follows. In Sec. II we review the wave packet dynamics theory
of Sundaram and Niu \cite{Niu}. In Sec. III we analyze the model for constant Berry curvature
obtaining in a physically clear way the reversal of the sign in the presence of smooth disorder.
In Sec. IV we apply the theory to the case of the Rashba model and show the non-trivial dependence
of the contributions from disorder scattering and the clean Berry phase contribution as a function
of the Fermi energy and in Sec. V we present our conclusions.

\section{The wave packet equations}
The motion of wave packets formed by Bloch states is governed by the following equations.\cite{Niu}
\begin{equation}
\frac{d}{dt} {\bf k}=e {\bf E} - {\bf \nabla}V({\bf r}) \label{dkx1}
\end{equation}%
\begin{equation}
\frac{d}{dt}{\bf r}=\frac{\partial \epsilon ({\bf k})}{\partial {\bf k}} - \frac{d {\bf k}}{dt}\times {\bf F}
\label{om1}
\end{equation}%
where $\epsilon (\bf{k})$ is the energy dispersion in the band, ${\bf E}$
 is the external electric field acting on a wave packet having the electric charge $e$, $V({\bf r})$ is the local potential
in the sample, for example the potential of impurities and 
 ${\bf F}$ is the Berry curvature of the Bloch band. For the two-dimentional motion only the out of plane component of $\bf{F}$ is nonzero 

\begin{equation}
F_{z}=2Im\left\langle \frac{\partial u^{s}}{\partial k_{y}}|\frac{%
\partial u^{s}}{\partial k_{x}}\right\rangle  \label{adi2}
\end{equation}%
here $\left\vert u^{s}\right\rangle$ is the Bloch state in the absence of the
electric field and impurities and $s$ is the index of the band.

The second term in (\ref{om1}) is responsible for the so called anomalous velocity. For the reader not familiar with the notion of the anomalous velocity
we provide the appendix with a simple example from classical physics that gives an intuitive explanation of the physical meaning of the anomalous
velocity in Rashba coupled 2DEG. For the rigorous theory we refer to the original papers \cite{Niu}.

The anomalous velocity is orthogonal to the direction of
the electric field \cite{{Lutt},{Wannier}} (which we chose to be along the $x$-axes). 
\begin{equation}
v_{y}^{(a)}=F_{z}\frac{dk_x}{dt}  \label{adi1}
\end{equation}%

The semiclassical equations (\ref{dkx1}) and (\ref{om1}) map the quantum mechanical problem to a classical one where particles have
the electric charge $e$ and move according to equations of the wave packet dynamics. 
We assume that at equilibrium the distribution of such classical particles is the same as the Dirac distribution of electrons in the sample.
This mapping to a classical system considerably simplifies the treatment of the problem both analytically and numerically. Analytically,
one can apply the classical Boltzmann equation approach to calculate the transport coefficients, numerically the molecular dynamics
 simulation of the motion of classical particles is simple  and may not be restricted to a small system size as, for example, in the case of a numerical diagonalization
of a quantum mechanical Hamiltonian. 

 Unfortunately equations  (\ref{dkx1}) and (\ref{om1}) are valid in the adiabatic limit only.
They cannot be applied to the case of scatterings on a short range delta-function like potential.
 Only scatterings on impurities, whose potential varies
appreciably only on distances much larger than the size of the wave packet can be calculated this way \cite{Niu}. 
In spite of this restriction of its applicability  the limit of a smooth potential is a very interesting one
to investigate the influence of disorder on the anomalous Hall effect.
In realistic applications a system with long range impurities 
is realized, for example, in the high mobility 2D
electron gas \cite{Aleiner} and the out of plane Zeeman field can be induced there by polarizing nuclear spins or by introducing additional magnetic impurities.

Recently another related effect, namely the intrinsic spin Hall effect, 
was introduced \cite{SHE}. A number of theoretical papers have explored the importance of the disorder.  The debates on this topic are ongoing  (see e.g. \cite{{Inoue},{Kentaro},{Halperin}}).
Understanding the AHE may shed light on the disorder role in the spin Hall effect.

\section{Constant Berry curvature}   
\subsection{Clean limit} 
In many recent applications the Berry curvature ${\bf F}$ strongly depends on the momentum of the wave packet $k$. \cite{{Jungwirth},{numerics1},{Dimi}}
However, 
it is instructive to consider first the one Bloch band 2D system with a constant Berry curvature $F_z$.

For a constant $F_z$ the anomalous velocity (\ref{adi1}) 
leads to the following contribution to the Hall current
from the unperturbed part of the distribution function in a single band at zero temperature:  
\begin{eqnarray}
&&J_{yx}^{(clean)}=e \int d^2 {\bf k} f_0({\bf k}) v_y^{(a)}=\nonumber\\
&=&e\int_0^{k_F}kdk \int _0^{2\pi} \frac{d\phi}{(2 \pi) ^2} eE_xF_z= \frac{e^2E_xF_z k_F^2}{4\pi}
\label{F1}
\end{eqnarray}   
where $ f_0({\bf k})$ is the equilibrium distribution function and $k_F$ is the Fermi momentum. The upper index $(clean)$ means that the quantity is calculated in the absence of disorder and the 
distribution function of the wave packets coincides with the one in equilibrium at $E_x=0$. 
  
  The case of a constant $F_z$ is realized in the conducting bands
  whose states are weakly hybridized with the states in the valent hole bands \cite{Nozieres} where the spin orbit coupling is allowed.
 If the kinetic energies of electrons in the conducting band are much smaller than the gap between conducting
  and valent zones, the Berry curvature due to such an induced spin orbit coupling can be considered as a constant for all conducting electrons.
  In modern applications it can be realized in some limiting situations, like in the case of R2DEG at a strong Zeeman field (see following sections). 
  In this section we show that for smooth impurity potentials 
  the wave packet approach easily reproduces the sign reversal of the Hall current when $F_z=const$. In comparison to previous works, however,
  our approach keeps the derivation  gauge invariant and hence is physically clear.

\subsection{Effects of disorder}  

  According to (\ref{F1}) if the distribution function is the same as in the equilibrium at zero external field  a Hall current appears
  in the external field  due to
  the anomalous velocity. However, in the steady state the distribution function is no longer $f_0 ( {\bf k})$. 
  In the electric field and on time scales much larger than 
  the scattering time particles diffuse with a constant velocity rather than accelerate as in the absolutely clean case. 
 The anomalous velocity and hence the clean contribution to the Hall current are proportional to the acceleration $\dot{k}_x$. Since in the 
 steady state the average acceleration is zero up to the first order in external electric field $E_x$ one can expect that disorder should 
 strongly influence the Hall current.
  
 A natural way to study the effect of disorder on the transport in our case is the classical Boltzmann equation. There is however a complication
 when both nonzero Berry curvature and  finite sizes of impurities must be considered. At a scattering on an impurity potential not only the 
 momentum but
 also the coordinate of a particle changes. Usually such a coordinate shift at the scattering is discarded since after averaging
 over many scatterings such random shifts cancel each other and only the changes of the momentum matters. However when the Berry curvature is nonzero
 the additional (anomalous) shift does not disappear after the averaging. To see this,
suppose that the term with the Berry curvature $F_z$ in Eq. (\ref{om1})
is small in comparison with the first one and calculate the corresponding correction
 to the shift of the particle during the scattering on an impurity. Integrating  (\ref{om1})
 over the time interval at which  a particle feels the impurity potential during a single scattering and treating the second term in (\ref{om1})
  as a small
perturbation one can find that, after the scattering, a particle makes an additional shift 
\begin{equation}
\Delta {\bf r}({\bf k}, {\bf k^{\prime }})=
{\bf z} \times ({\bf k}^{'}-{\bf k}) F_{z} + \int_{t_1}^{t_2} 
\frac{\partial \epsilon}{\partial {\bf k}} dt 
\label{zse}
\end{equation}
here ${\bf k}^{\prime }$ and ${\bf k}$ are momentums respectively after and before the scattering, $t_1$ and $t_2$ are times of entering and leaving
the impurity in a semiclassical picture.
The second term in the rhs. of (\ref{zse}) is just the shift due to the normal velocity. To first order in $F_z$ and 
${\bf \nabla} V$ it is not affected by the Berry
phase and hence is averaged to zero after many scatterings; therefore we will disregard it in our
future discussion. 

The particle's displacement due to the first term in (\ref{zse}) is due to the 
anomalous velocity.
This shift does not depend explicitly on the details of the impurity potential and does not have the chiral symmetry. 
There are two main rather distinct effects due to the appearance of this anomalous shift. 
The first effect, the so called side-jump,
is that the $y$-component of this shift does not cancel after the averaging over many scatterings and thus contributes to the drift velocity perpendicular to the electric field, i.e. to the Hall current.
We will focus on this effect in the following subsection. 
The second effect is that when a scattering takes place in the presence of an external electric
field there is a change in the potential energy upon a scattering given by 
\begin{equation}
\Delta U=-eE_x\Delta x  \label{b1}
\end{equation}%
where $\Delta x$ is the  shift along the external electric field.\cite{{Berger1},{Berger2},{Berger3}}
Both of these effects, as shown below and in the next subsection, give the same contribution as
Eq. (\ref{F1}) but with an opposite sign in the particular case of a momentum-independent Berry curvature.
Because both contributions are linear in $E_x$ we are able to consider them separetly when considering
the linear response of the system.

Let us focus first on the second effect. Here only the anomalous part of this 
shift is important for Hall current calculations
\begin{equation}
\Delta x=-F_{z}(k_{y}^{\prime }-k_{y})  \label{b2}
\end{equation}
and the effect of the normal part of the shift only renormalizes the diagonal current response to $E_x$. Since the
total energy remains the same after an elastic scattering, the kinetic energy
must change by the amount
\begin{equation}
\Delta \epsilon ({\bf k},{\bf k^{\prime}})  = \epsilon ({\bf k^{\prime }})-\epsilon ({\bf k})=eE_x\Delta x
\label{b55}
\end{equation}

This effect leads to an instability of the initial equilibrium distribution
function $f_{0}({\bf k})=f_0(\epsilon({\bf k}))$ due to scatterings. 
The Boltzmann equation  is given by 
\begin{equation}
\frac{\partial f}{\partial t}=-\sum_{\bf k^{\prime }}\omega ({\bf k,k^{\prime
}})[ f \left(\epsilon ({\bf k}) \right) -f\left(\epsilon ({\bf k^{\prime }})\right) ]  \label{b3}
\end{equation}%
where $\omega ({\bf k,k^{\prime
}})$ is the scattering rate.
Kinetic energies before and after a scattering do not coincide but the difference between them is small.
Since we are seeking a contribution linear in $E_x$ we can substitute $f$ by $f_{0}$ which 
depends only on the kinetic energy $\epsilon ({\bf k})$ so we can expand
\begin{equation}
\begin{array}{l}
f_{0}(\epsilon ({\bf k}))-f_{0}(\epsilon ({\bf k^{\prime
}}))= \\ 
\\ 
=\frac{-df_{0}}{d\epsilon }(\epsilon ({\bf k^{\prime }})-\epsilon ({\bf k}))=-\frac{%
df_{0}}{d\epsilon }eE_xF_{z}(k_{y}-k_{y}^{\prime })%
\end{array}
\label{b4}
\end{equation}%
This yields 
\begin{equation}
\frac{\partial f}{\partial t}=-\sum_{\bf k^{\prime }}\omega ({\bf k,k^{\prime }})\frac{%
-df_{0}}{d\epsilon }eEF_{z}(k_{y}-k_{y}^{\prime })\neq 0  \label{b5}
\end{equation}%
The distribution function will relax until a contribution to it
compensates this relaxation in the stationary limit.

To find the new equilibrium distribution one should substitute $f({\bf k}) = f_0 (\epsilon  ({\bf k})) + g_{E}({\bf k})$ into the rhs. of equation
(\ref{b3}).
  To make $\partial f/\partial t=0$  in  equation  (\ref{b5}) the correction $g_{E}({\bf k})$ must be the following
  \begin{equation}
g_{E}({\bf k})=-\frac{-df_{0}}{d\epsilon }eE_xk_{y}F_{z}
  \label{b6}
\end{equation}%
This contribution is not symmetric in the $y$-direction. This means that already
the normal (i.e. usual $\partial \epsilon /\partial {\bf k}$) velocity can contribute to the Hall current in the stationary state:

\begin{eqnarray}
J_{yx}^{normal}&=&e\int \frac{d^{2}{\bf k}}{(2\pi )^{2}}g_E({\bf k})(v_y^{(normal)})\nonumber\\
&=&-\frac{e^2E_x F_z}{ (4\pi )}k_{F}^2
\label{jnos}
\end{eqnarray}
which has the opposite sign and is exactly the same in the absolute 
magnitude as the anomalous velocity contribution in  the clean limit given by Eq.(\ref{F1}).

\subsection{The side jump}

The anomalous change of energy after the scattering is not the only 
important effect of the anomalous shift (\ref{b1}).  
The anomalous shift during the scatterings has generally a 
component perpendicular to the direction of the electric field.
\begin{equation}
\Delta y({\bf k}, {\bf k^{\prime }})=F_{z}(k_{x}^{\prime }-k_{x})
\label{shift22}
\end{equation}
It does not contribute to the change of energy during the scattering but it shifts a particle along the $y$-axes. If such shifts do not
compensate each other after many scatterings, they should contribute to the total Hall current. This phenomenon is known in the theory of the
extrinsic Hall effect as the side jump. If scatterings happen with the
rate $\omega ({\bf k,k^{\prime
}})$ in average the particle moving with the momentum
${\bf k}$ also acquires the anomalous drift velocity perpendicular to the electric field 
\begin{equation}
\left\langle v_{y}^{(sj)}({\bf k})\right\rangle _{imp}=\sum_{\bf k^{\prime }}\omega
({\bf k}, {\bf k^{\prime }})\Delta y({\bf k},{\bf k^{\prime }}) =-F_z k_{x}/\tau _{||} 
 \label{sum}
\end{equation}
where 
\begin{equation}
1/\tau _{||}= \sum_{{\bf k^{\prime }}} \omega ({\bf k,k^{\prime }})(1-\cos ({\bf k,k^{\prime}})) %
\label{tau1}
\end{equation}
and "$(sj)$" marks the side jump contribution to the physical quantity.

In the equilibrium the side jump does not lead to the Hall current because the anomalous velocity (\ref{sum}) changes 
the sign under the transformation $k_x \rightarrow -k_x$. 
and the distribution function in the equilibrium is invariant under this transformation. 
However, when  the electric field
is applied the non-equilibrium correction to the distribution function 
appears which has no such a symmetry under the momentum reflection.
 
This correction to the distribution function
can be derived by means of the standard approach to the Boltzmann equation.
 Up to the first order in the electric field the standard Boltzmann equation for one band, ignoring
the asymmetric contribution considered in the previous subsection, is%
\begin{equation}
-eE_x v_{x}^{(normal)}(-\frac{\partial f_{0}}{\partial \epsilon })=-\sum_{\bf k^{\prime
}}\omega ({\bf k,k^{\prime }})(f({\bf k})-f({\bf k^{\prime }}))
\label{be2}
\end{equation}
where $f_0({\bf k})$ is the equilibrium distribution function. 
Equation (\ref{be2}) has a solution $f({\bf k})=f_{0}({\bf k}) +g({\bf k})$ where to the first order in the electric field
\begin{equation}
g({\bf k})=eE_x\tau _{||}(-\frac{\partial f_{0}}{\partial \epsilon })v_{x}^{(normal)}
\label{g1}
\end{equation}%
$v_{x}^{(normal)}$ is the normal velocity along the electric field, which in our case is
\begin{equation}
v_{x}^{(normal)}=\frac{\partial \epsilon (k)}{\partial k_{x}}=bk_{x}
\end{equation}
here we assumed that the conducting band has a trivial dispersion $\epsilon ({\bf k})=bk^2/2$ where $b=1/m_e$.
The side-jump contribution of impurities to the anomalous velocity leads to the Hall
current
\begin{eqnarray}
J_{yx}^{(sj) } &=&e\int \left\langle v_{y}^{(sj)}({\bf k})\right \rangle _{imp}g({\bf k })%
\frac{kdkd\phi }{(2\pi )^{2}}
\nonumber\\&=&-\frac{e^2E_xF_z}{(2\pi)^2}
\int_0^{2\pi} d\phi \cos ^{2}(\phi )\int d\epsilon_k k^2\delta (\epsilon_F
-\epsilon _{k}) \nonumber\\  
&=&-\frac{e^2E_x F_z}{ (4\pi )}k_{F}^2\label{fin2}
\end{eqnarray}
As in the previous impurity contribution, the Hall current due to the side jump has the opposite sign and the same magnitude as the one in the clean limit.
The side jump current appears because the anomalous velocity is proportional to the acceleration
 of the particle but in the stationary state the average acceleration should be zero 
 (up to the terms of  the first order in the electric field), namely, while the external electric field accelerates the wave packet during its motion between 
 impurities, the impurity potential generally decelerates it. During such  a deceleration the wave packet has the anomalous velocity with the opposite sign
 and hence the side jump contribution should compensate or at least decrease the pure limit result.
The described picture is of course oversimplified, which will be clear from the discussion of a momentum dependent $F_z$.
In reality, the external electric field equally
accelerates all electrons in the Fermi sea but effective deceleration due to impurity scatterings is, in average, seen only by the electrons near
the Fermi level because there the distribution function acquires a correction required to compensate the acceleration by the external field.
Hence the side jump contribution is the property of the Fermi surface while the clean limit Berry phase contribution appears from all electrons even deep in the Fermi sea.

\subsection{Total current and numerical check}

The total Hall current (not including effects of skew scatterings due to asymmetry of the collision term kernel)
is the sum of the clean limit contribution,
of the side jump and of the normal velocity contributions. For the constant $F_z$ we find
\begin{equation}
J_{yx}^{(total)}= J_{yx}^{(clean)}+J_{yx}^{(sj)}+J_{yx}^{(normal)}=-\frac{e^2E_x F_z}{ (4\pi )}k_{F}^2
\label{cfin}
\end{equation}
As it was predicted in the former literature which focused on models with a constant $F_z$, 
 the total current has the same magnitude but with the opposite sign as the one in the 
absolutely clean system. Our derivation however is
considerably simpler and is straightforward to generalize to a case with a 
more complicated band structure such as Rashba coupled 2DEG.

\begin{figure}
\includegraphics[width=3.2 in]{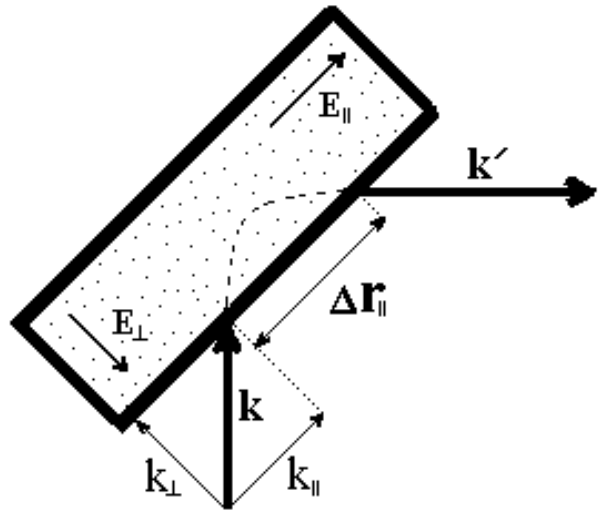}
\centering
\caption{Scattering of a particle on impurity. The impurity is assumed to have the shape of a long stripe with linearly growing potential inside. }
\label{Wall}
\end{figure}
To confirm the analytical result (\ref{cfin}) numerically we simulate the motion of 
particles according to equations of motions (\ref{dkx1}) and (\ref{om1}). 
Initially all particles were prepared uniformly distributed over the 
momentum space having  the absolute value of the momentum less
than a specified Fermi momentum. The action of an impurity was simulated by the potential of a wall of a finite thickness
 with a very fast linearly growing potential inside the wall. The wall is sufficiently long to disregard
  the effects on its edges. This type of impurity does not allow the skew scattering mechanism to appear because for every scattering
   on the wall that
  changes momentum from ${\bf k}$ to ${\bf k^{\prime}}$ there is a process when a particle hits the same wall from another side and scatters from
  ${\bf k^{\prime}}$ into ${\bf k}$. Thus both processes have equal probability and $\omega({\bf k,k^{\prime}})=\omega({\bf k^{\prime},k})$. 
 Fig.\ref{Wall} demonstrates a scattering on such an impurity. We would like to point to the analogy of the scattering in Fig.\ref{Wall}
 with Goos-H\"anchen's and with Fedorov's shifts in geometrical optics \cite{optics}.

One can consider the effect of the wall on a wave packet as the one
 due to an electric field $E_{\perp}$ acting on it only inside the wall in the direction perpendicular to the wall.
Although the potential in the wall can be strong in comparison with the external electric field, we suppose that the motion inside the wall is still
governed by
the equations of the wave packet dynamics. Such a scattering problem can be solved exactly. This solution shows that for the given type of
 scatterings
the theory of the anomalous shift is valid even if we do not assume that the anomalous velocity term in wave packet equations is small. 
This is useful because the bigger anomalous velocity is needed to accelerate the numerical calculations.  

 Let components of the momentum of the particle incident on the wall be 
$k_{||}$ and $k_{\perp}$ parallel and perpendicular to the wall respectively.  
In the presence of an additional external electric field 
the total force has generally also a nonzero component parallel to the wall $eE_{||}$ where $E_{||}$ is the 
projection of the external electric field on the direction along the wall. Expressions for the final momentum and
 coordinates right after the scattering
are strongly simplified in the limit of large $E_{\perp}$, so that all terms $O(1/E_{\perp})$ can
be dropped. In this limit
 the scattering time is vanishing and in (\ref{om1}) one can disregard the first term in comparison with the second one since
 $\dot{k}_{\perp} \sim E_{\perp}$ is large. Dropping the term with $\partial \epsilon/\partial k_{\perp}$ the equation (\ref{om1}) can be
readily integrated over the time of the scattering leading to the relations $\Delta r_{||}=eF_z(k_{\perp}-k_{\perp}^{\prime})$ and $k_{||}^{\prime}=k_{||}$. 
Solving them together with the energy conservation equation $\epsilon ({\bf k})+eE_{||}r_{||}=\epsilon ({\bf k^{\prime}})$ 
we arrive at following expressions for the momentum of the outgoing particle
\begin{equation}
\begin{array}{l}
k_{||}^{\prime}=k_{||}\\
k_{\perp}^{\prime}=-k_{\perp}-E_{||}F_z
\end{array}
\label{kkk1}
\end{equation}
 In addition, the scattered particle appears
in a point of the interface shifted in comparison to the incident point by the amount. 
\begin{equation}
\Delta r_{||}=2k_{\perp}F_z+E_{||}F_z^2
\label{kkk2}
\end{equation}
The shift $r_{||}$ is exactly the anomalous shift in the impurity potential. Note also that, as it follows from
(\ref{kkk1}), when a scattering happens in the additional external
electric field with nonzero projection to the direction of this shift $E_{||}$, the absolute magnitude of the momentum changes, which leads to 
the instability of the equilibrium distribution function, discussed in previous sections. 
The scattering time in this
limit is small enough to be disregarded so numerically such a scattering can be easily simulated by 
choosing the coordinate system related to the wall orientation and updating the momentums and coordinates according to
$(k_{||},k_{\perp}, r_{||}, r_{\perp}) \rightarrow (k_{||}^{\prime}, k_{\perp}^{\prime}, r_{||}+\Delta r_{||},r_{\perp})$,  $r_{\perp}$ is 
the same for the incident and the out going particle. In between scatterings equations of free motion in the external electric field are also trivially 
solvable. In our simulations we assumed the "noncrossing approximation", namely we suppose 
that every scattering happens on a different impurity. 
The algorithm consists of two parts. First we generate randomly the distance $L$ to the next impurity.
We suppose that between impurities particles move only under the action of the electric field. The integration of wave packet equations
 leads to the following
trajectories
\begin{equation}
\begin{array}{l}
k_x^{\prime}=k_x+E_xt\\
k_y^{\prime}=k_y\\
x^{\prime}=x+k_xt+\frac{E_xt^2}{2}\\
y^{\prime}=y+(k_y+F_zE_x)t
\end{array}
\label{traj}
\end{equation}
The time $t$ of the motion to the next impurity was estimated from the equation $((x^{\prime}-x)^2+(y^{\prime}-y)^2))^{1/2}=L$.
Solving this relation with the solution (\ref{traj}) we find the 
momentum and coordinates of the particle before
the new scattering. In the second step we randomly generate the orientation angle of the impurity, switch into the related coordinate system
 and update  particle's coordinates and the momentum according to
(\ref{kkk1}) and (\ref{kkk2}). Then we return to the initial coordinate system and repeat the circle.
 We repeat this circle sufficiently many times to allow the distribution to relax to the 
stationary one (usually a few scatterings is enough),
however the total evolution time was small enough in order to avoid strong heating of the system. In our simulations, 
every particle makes about $10^2-10^3$ scatterings during the whole time. To prevent strong heating the electric field is chosen sufficiently
small $E_xl/E_F \sim 10^{-4}$ where $l$ is the typical scattering length and $E_F$ is the Fermi energy.
The total current can be derived as the sum of total displacements of all particles divided by the evolution time.  

\begin{figure}
\includegraphics[width=3.2 in]{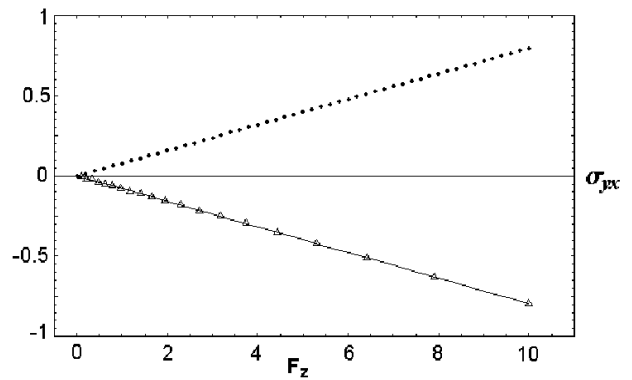}
\centering
\caption{Hall conductivity vs strength of the Berry curvature in the model with momentum independent $F_z$. Solid line is the theoretical prediction that includes impurity
scatterings and dotted line is the prediction of the clean limit contribution. Triangles show results of the numerical simulations.
Simulations were performed in units $e=\hbar = m=1$ and  $k_F=1$.}
\label{fig1}
\end{figure}
In Fig.\ref{fig1} we compare the analytical result (\ref{cfin}) 
with our numerical simulations for the fixed Fermi energy but different
strengths of the spin orbit coupling, and hence $F_z$. Both analytical and numerical 
results are in excellent agreement with each other. 
This result survives at an arbitrary magnitude of the Berry curvature.

\begin{figure}[h]
\includegraphics[width=2.4 in]{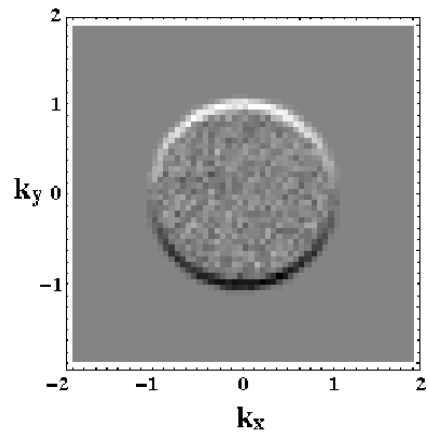}
\centering
\caption{Distribution of particles over the momentum space in the stationary state with electric field applied along the $y-$axes at zero Berry
curvature. Parameters are as follows. $k_F=1$, $E_y=0.003$, $E_x=0$, $F_z=0$. Average distance between impurities is $l \sim 0.5$ in units
$m=\hbar=e=1$.}
\label{dist1}
\end{figure}
\begin{figure}[h]
\includegraphics[width=2.4 in]{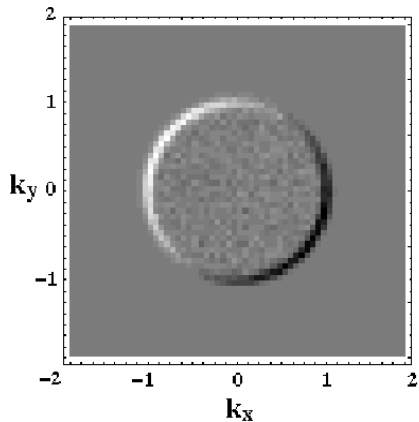}
\centering
\caption{Distribution of particles over the momentum space in the stationary state with electric field applied along the $y-$axes at 
the non-zero Berry
curvature. Parameters are as follows. $k_F=1$, $E_y=0.003$, $E_x=0$, $F_z=-1$. Average distance between impurities is $l \sim 0.5$.}
\label{dist2}
\end{figure}

In Fig.\ref{dist1} and Fig.\ref{dist2} we visualized the stationary distribution function of particles in the momentum space, calculated numerically
for the system placed in  the uniform electric field.  Brighter areas represent
higher densities. To increase the contrast, we subtracted the initial equilibrium distribution (i.e. without electric field)
 from the calculated one.  In  
 Fig.\ref{dist1} the Berry curvature is set to zero. In this case the correction to the equilibrium distribution is symmetric along the 
 electric field. Fig.\ref{dist2} shows that when the Berry curvature becomes nonzero, the distribution function
 acquires the additional asymmetric contribution, which obviously deposits to the Hall current.

\subsection{Analogy between extrinsic and intrinsic AHE}
As we showed above, in the AHE induced by Berry curvature there is a possibility of disorder
effects similar to previously studied  extrinsic contributions to Hall current due to gradient terms of
the disorder potential. 
In this subsection we discuss how this relation is encoded in the wave packet equations (\ref{dkx1}), (\ref{om1}). We consider here only the case 
of a constant Berry 
curvature. The wave packet dynamic equations do not arise from a particular Hamiltonian in a sense that coordinates and momentums of a wave packet
are not canonical variables. 
Nevertheless, it is possible to make them arise from a Hamiltonian by adding terms of 
higher order in the potential gradient. We remind the reader  that equations (\ref{dkx1}) and (\ref{om1}) 
were derived in the approximation of a smooth potential so that 
its higher gradients could be discarded. 
Therefore such a procedure should not change the physics at least to the order of the approximation of the 
whole theory. Consider the following system
\begin{equation}
\frac{d}{dt} {\bf k}=e {\bf E} - {\bf \nabla}V({\bf r}) -  {\bf \nabla} ({\bf k}\times {\bf \nabla}V({\bf r})) \bf{F} \label{dkx11}
\end{equation}%
\begin{equation}
\frac{d}{dt} {\bf r}=\frac{\partial \epsilon ({\bf k})}{\partial {\bf k}} - (e {\bf E} - {\bf \nabla}V({\bf r}))\times {\bf F}
\label{om11}
\end{equation}
Except for one term of the second order in gradients of the potential the system is equivalent to the wave packet equations
(\ref{dkx1}),  (\ref{om1}). 
The new system (\ref{dkx11}), (\ref{om11}) already corresponds to the classical Hamiltonian evolution with the following Hamiltonian

\begin{equation}
H({\bf k, r})=\epsilon({\bf k}) - {\bf F} ({\bf k}\times (e{\bf E}-{\bf \nabla} V({\bf r}))+V({\bf r})-e{\bf E}{\bf r}
\label{hhhhh}
\end{equation}
This Hamiltonian has the same structure as in typical models of the extrinsic AHE \cite{{Berger1},{Berger2},{Berger3}} and the Berry
curvature now plays the role of an effective spin orbit coupling constant. As we mentioned,  the
main contributions to the extrinsic Hall current  have been proved to be the side jump and the skew scattering.
In our previous calculations, including numerics, we ignored the
skew scattering mechanism, 
nevertheless  the skew scattering  can be calculated by same 
weave packet techniques because the second term in the rhs. of (\ref{hhhhh})  is responsible 
for both skew scattering and the side jump contributions in the extrinsic AHE. 
Note also that the skew scattering current will depend on the Berry 
curvature and thus will also have a geometric interpretation. The fact that the side jump is the only
other surviving contribution in the extrinsic AHE is in agreement with our previous
findings.

\section{Rashba 2DEG with exchange coupling}

Recently studied systems with AHE such as DMS and Rashba 2DEG have Berry curvature that strongly 
depends on momentum of the wave packet \cite{Dimi}. When $F_z$ is no longer a constant the simple arguments leading to canceling of some terms
may not work. One
can notice that when an electron decelerates it has in general different momentum
than when it accelerates,  in other words the uniform electric field
accelerates all electrons down to the bottom of the Fermi sea but impurities
produce nonzero deceleration in average only for electrons near the Fermi
surface. The Berry phase acquired by accelerating
electrons depends now not only on the acceleration but on the momentum itself
therefore contributions from impurities and from uniform electric field not
necessarily cancel each other. 

The Hamiltonian of R2DEG with the electric field along the $x$-direction and
the exchange field in the $z$-direction is

\begin{equation}
H=H_{0}+H_{SO}+H_{exch}-eE_{x}x+V_{imp}  \label{hs0}
\end{equation}
\begin{equation}
H_{0}=\frac{bk^{2}}{2}\sigma _{0},\,\,\, b=1/m  \label{h0}
\end{equation}
\begin{equation}
H_{SO}=\lambda (k_{y}\sigma _{x}-k_{x}\sigma _{y})  \label{hso1}
\end{equation}
\begin{equation}
H_{exch}=h\sigma _{z}  \label{exch1}
\end{equation}
where $V_{imp}$ is the impurity potential.

Diagonalizing the unperturbed part of the Hamiltonian we find the energy dispersion

\begin{equation}
\epsilon_{\pm} (k) = bk^2 \pm \sqrt{(\lambda k)^2+h^2}
\label{disp}
\end{equation}
There are two bands, the minor band has the plus sign in the above expression. Fermi momentums can be derived by inversion of the equations
(\ref{disp}). We denote them as $k_{F-}$ and $k_{F+}$ respectively for the major and minor bands.
The Berry curvature is different for different bands 
\begin{equation}
F_{z}^{s}=2Im\left\langle \frac{\partial u^{s}}{\partial k_{y}}|\frac{%
\partial u^{s}}{\partial k_{x}}\right\rangle  \label{adi2}
\end{equation}%
here $\left\vert u^{s}\right\rangle$ is the Bloch state in the absence of
electric field and impurities and $s=\pm$ is the index of the band: the plus is for the minor and the minus is for the major one.
In the case of the Hamiltonian (\ref{hs0}) the Berry curvature is \cite{Dimi}, (see also Appendix for an alternative derivation).
\begin{equation}
F_z^{s}(k)=-s\frac{h \lambda^2}{2[(\lambda k)^{2}+h^2 ]^{3/2}}
\label{bc1}
\end{equation}
at  $\lambda k<<h$ this gives $F_z^{s}(k)\approx -s \lambda^2/(2h^2)=const$ and we reduce the problem to the one considered in previous sections.
In the opposite limit at $\lambda k>>h$ we find
\begin{equation}
F_z^{s}(k)\approx -\frac{sh}{2\lambda k^{3}}
\end{equation}


The anomalous velocity has the same magnitude but different signs for different bands so filled states
with the same $\mathbf{k}$ from different bands will compensate each other in calculations of the clean contribution and
to find the Hall current we should only integrate over uncompensated states of the  major band. The Hall
current from the unperturbed electron distribution function in the clean case  is
\begin{eqnarray}
&&j_{yx}^{(clean)}=e\int\limits_{0}^{2\pi }d\phi \int\limits_{k_{F+}}^{k_{F-}}\frac{kdk}{(2\pi )^{2}}
(eE_xF_z^{-}(k))\nonumber\\&=&e\int\limits_{0}^{2\pi }d\phi
\int\limits_{k_{F+}}^{k_{F-}}\frac{kdk}{(2\pi )^{2}}\frac{eE_xh \lambda ^2}{2[(\lambda k)^{2}+h^2]^{3/2}%
}\nonumber\\
&=&\frac{e^2E_x}{2(2 \pi)} \left( \frac{1}{\sqrt{1+(k_{F+} \lambda /h)^2}}\right.\nonumber\\&&\left.
-\frac{1}{\sqrt{1+(k_{F-} \lambda /h)^2}} \right)
\label{fin1}
\end{eqnarray}
in the case $E_{F}>>  \lambda k_{F\pm}>>h$ the expression (\ref{fin1}) is simplified
\begin{eqnarray}
j_{yx}^{(clean)}&\approx&
\left\{ \lambda k\ll E_{F}\approx \frac{k_{F}^{2}}{2m}\right\} =\frac{%
e^{2}E_xh(k_{F-}-k_{F+})}{(2\pi )2\lambda k^{2}}\nonumber\\&=&\left\{
(k_{F-}-k_{F+}) \approx 2\lambda m\right\} =\frac{e^{2}E_xh}{4\pi E_{F}}
\label{fin11}
\end{eqnarray}%
note that here $k_{F+}$ is the Fermi momentum of the minor band.
The above formulas are valid when there are electrons in both bands. At a sufficiently low Fermi level $(E_F<h,\lambda k_{F-})$ the minor band becomes 
depleted and the transport is only in the major band. In this case the clean limit Hall current is
\begin{equation}
j_{yx}^{(clean)} 
=\frac{e^2E_x}{2 (2 \pi)} \left(1 - \frac{1}{\sqrt{1+(k_{F-} \lambda /h)^2}} \right)
\label{fin2}
\end{equation}
We address next the effects of impurity scattering within the formalism that we have developed.

As in the case of the constant Berry curvature, the
 non-equilibrium contribution to the distribution function is the sum of the usual one $g^s ({\bf k})$,
  which appears to compensate the electron acceleration
\begin{equation}
g^s ({\bf k})=eE_x\tau _{||}(-\frac{\partial f_{0}}{\partial \epsilon })v_{s,x}^{(normal)}
\label{g11}
\end{equation}%
 and of the anomalous one $g_{E}^s({\bf k})$
   due to the shift of the kinetic energy of the scattered particle. We suppose here 
   that the impurity potential is weak, namely $V({\bf r})<<E_F, \lambda k_F$ and disregard the variation 
of the Berry curvature near a wave packet during a scattering because the absolute magnitude of the momentum does not change appreciably.
The derivation of the anomalous correction is analogous to (\ref{b3})-(\ref{b6}).
  \begin{equation}
g_{E}^s({\bf k})=-\frac{-df_{0}}{d\epsilon }eE_xk_{y}F_{z}^s
  \label{bb6}
\end{equation}%
The normal velocity of the wave packet is
\begin{equation}
v_{s,x}^{(normal)}=\frac{\partial \epsilon_s (k)}{\partial k_{x}}  \label{vnx}
\end{equation}

The expression for the drift velocity due to the side jumps along the $y$-axes is the same as in (\ref{sum}).
The side jump current is
\begin{eqnarray}
&&j_{yx}^{(sj) }=\sum_{s=\pm}e\int \langle v_{s,y}^{(sj)} (\mathbf{k}) \rangle _{dis} g^s(\mathbf{k)%
}\frac{kdkd\phi }{(2\pi )^{2}}\nonumber\\&=&\sum_{s=\pm}
\int d\phi \frac{\cos ^{2}(\varphi )}{(2\pi )^{2}}\int d\epsilon_{s,k}  e^2 E_xF_z^s (k)k^2 \delta (E_F
-\epsilon _{s,k})= \label{fin22} \nonumber\\ 
&=&\frac{he^{2}E_x \lambda^2}{2 (4\pi )}\left(\frac{k_{F+}^2}{[(\lambda k_{F+})^{2}+h^2]^{3/2}}\right.
\nonumber\\&&\left.
-\frac{k_{F-}^2}{[(\lambda k_{F-})^{2}+h^2]^{3/2}} 
\right) 
\end{eqnarray}
At $E_F>>\lambda k_F>>h$ this expression can be simplified
\begin{equation}
j_{yx}^{(sj) } \approx \frac{he^{2}E_x}{8\pi E_F}
\label{alsdfj}
\end{equation}
If $E_F<h$ only the major band survives and
\begin{equation}
j_{yx}^{(sj) } = \frac{-he^{2}E_x \lambda^2}{2 (4\pi )} \frac{k_{F-}^2}{[(\lambda k_{F-})^{2}+h^2]^{3/2}}
\label{fin222}
\end{equation}

The normal current contribution is
\begin{equation}
\begin{array}{l}
j_{yx}^{(normal)}=\sum_{s=\pm}e\int \frac{d^{2}{\bf k}}{(2\pi )^{2}}%
v_{s,y}^{(normal)}g_{E}^s({\bf k})= \\ 
\\ 
=\sum_{s=\pm}
\int d\phi \frac{\sin ^{2}(\varphi )}{(2\pi )^{2}}\int d\epsilon_{s,k}  e^2 E_x F_z^s (k)k^2 \delta (E_F
-\epsilon _{s,k})
\end{array}
\label{bb7}
\end{equation}%
Comparing the last expression with the one for the side jump current (\ref{fin22}) we find that they are different only by an exchange of 
$\cos(\varphi)$ and $\sin(\varphi)$ under the integral over the angle. Thus we arrive at the general result 
\begin{equation}
j_{yx}^{(normal)}=j_{yx}^{(sj) }
\label{jisj}
\end{equation}
Finally,
\begin{equation}
j_{yx}^{(total)}=j_{yx}^{(clean)}+j_{yx}^{(sj) }+j_{yx}^{(normal)}
\label{totl}
\end{equation}

As it is seen from (\ref{alsdfj}), (\ref{fin11}) the side jump contribution may not change the sign in comparison with the 
clean contribution for sufficiently large
Fermi energy. This is very distinct
from the case of the constant Berry curvature where such a
contribution exactly cancels the pure one. At the limit $E_F>>\lambda k_F>>h$ we find for the total current
\begin{equation}
j_{yx}^{(total)}\approx \frac{he^{2}E_x}{2\pi E_F}
\label{totlim1}
\end{equation}
The total Hall current not only has the same sign, as the clean limit prediction, but also increases due to scatterings. Note also, that 
up to a coefficient of the order unity, in this limit, the total current has the same dependence on various parameters.

\begin{figure}
\includegraphics[width=3.0 in]{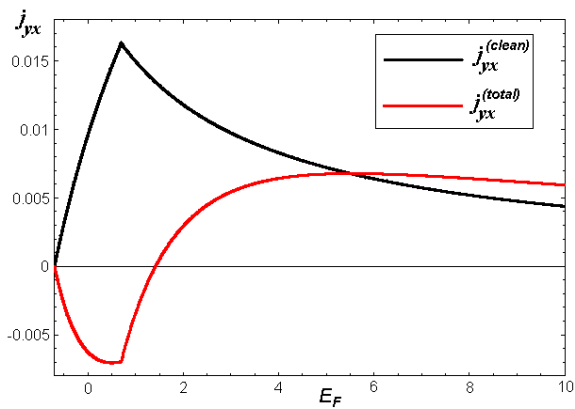}
\caption{(Color online) Hall conductivity vs the Fermi energy in the R2DEG. Black line is the pure limit Berry phase prediction. The red curve shows the prediction
of the theory that includes impurity
scatterings (\ref{fin22}), (\ref{jisj}).
Data are given in units $e=\hbar = m=1$ for the case $h=0.7$, $\lambda = 0.4$. Sharp change in the behavior above $E_F=0.7$ is due to the 
appearance of electrons in the minor band. }
\label{Rcomp}
\end{figure}

In Fig.\ref{Rcomp} we compare the theoretical prediction of Eqs. (\ref{totl}) with the prediction of the pure Berry phase contribution (\ref{fin1}).

\section{Discussion and Conclusion}

We calculated the anomalous Hall current in the presence of a smooth disorder potential by 
combining the wave packet dynamic equations with  the classical Boltzmann equation.
 For the case of a constant 
Berry curvature our results are in agreement with some known predictions of the very earliest works on AHE. The gauge invariance of our approach
clarifies the physical meaning of various disorder 
contributions to the Hall current and reveales the relation between
the extrinsic, related to the gradients of the disorder potential,
 and Berry curvature effects in the presence of disorder. 

We have shown that early results, like that of the Luttinger's sign reversal prediction,
 cannot be used directly for new applications such as diluted magnetic semiconductors, where the 
Berry curvature is strongly momentum dependent. 

Our prediction for the Hall current Fig.\ref{Rcomp} is distinct from the clean limit (or the pure Berry phase) one. Thus the total Hall current changes the sign when the Fermi level increases. 
In contrast, the clean limit prediction 
for the models considered remains of the same sign.
The filled states in both bands
with the same momentum compensate each other, 
hence for the clean case the band having more filled states always wins. In contrast, 
impurity contributions are sensitive not only to the 
density of states near the Fermi level, but also to the 
magnitude of the Berry curvature near the Fermi level. 

The change of sign of the Hall current as a function of chemical potential originates from the competition between two bands.
The minor band has the smaller number of filled states than the major one but it also has lower $k_F$ and hence the stronger $F_z(k_F)$.
According to our calculations,
such a change of sign should generally happen when the 
chemical potential is close to the depletion point of the minor band. 
When the minor band is totally 
depleted we always observe the reversal of the sign of the Hall current in comparison to the clean limit contribution. 
We also note that such a competition should be applicable to the skew scattering mechanism as well. This follows from the fact that the
strength of the skew scattering should be proportional to 
the Berry curvature at the Fermi surface.

Our results can be generalized to DMS and although in this case the complicated 
behaviour of the bands, as compared to the R2DEG, may change the quantitative behavior
there is still a competition in DMS 
of Hall currents from major and minor bands and Berry curvature is also usually decreasing 
when momentum is increasing. However the intricate dependence of the clean limit AHE observed
in these systems versus the simple dependence observed in the R2DEG will require a careful analysis.

Although we have verified most our predictions by numerical simulations, 
the correspondence between the wave packet approach and the evolution of a 
true quantum system with disorder remains to be investigated. 
An effect not taken in to account in our calculations is
 that the spin density in the wave packet polarizes when the wave packet
is accelerated \cite{Dimi2}. This effect is crucial for 
the Spin Hall Effect and its importance for calculations by the Boltzmann equation
should be understood.

The final theory of disorder in the AHE can be achieved 
only by purely quantum mechanical calculations for example, based on the  Keldysh technique. 
At present new approaches to systems with Berry curvature have 
emerged \cite{Haldane}, as well as new mechanisms to generate the anomalous Hall current
\cite{Bruno-rods}, but we
believe that the wave packet approach combined with the classical 
Boltzmann equation is worth studying because it provides the simplest,
to our knowledge, demonstration of the related physics and the wave packet equations 
themselves have been well justified by the quantum theory. 

\noindent
 {\it Note added}. After completion of this work we became aware of 
the related effort by V. K. Dugaev {\it et al.} \cite{Dugaev} of a more quantum mechanical approach
to the problem with similar qualitative conclusions. Our approach
however is quite different and in combination with this work may shed further light on the 
physical interpretation of the effect.

\noindent
 {\it Acknowledgments}. The authors are greatful for insightful and useful discussions with 
A. H. MacDonald and V. L. Pokrovsky. This work was supported by Welch Foundation, by DOE 
grant DE-FG03-02ER45958 and by NSF under the grant DMR-0115947. 

\appendix
\section{Classical interpretation of the wave packet dynamics in a Rashba 2DEG}

To make the semiclassical result more transparent we show how the anomalous velocity arising from
the Berry curvature can appear in a purely classical system. This classical anomalous velocity originates from 
the non-adiabatic contributions to the equations of motion in the linear response regime and are not present
when a simpler adiabatic approximation is considered.

We consider the motion of a classical particle 
having the electric charge $e$ and the classical spin ${\bf S}$ attached to it. Choose the Hamiltonian to be
\begin{equation}
H=\frac{k^2}{2m}+2\lambda(k_y S_x - k_x S_y) - eE_x x +2hS_z
\label{hcl1}
\end{equation}
The analogy with the Rashba Hamiltonian (\ref{hs0}) is obvious. In the classical problem we substitute operators by corresponding classical variables.
We will suppose also that $|S|=1/2$ as for true electrons.
The Hamilton equations for the evolution of coordinates $(x,y)$ and momentum $(k_x,k_y)$ and 
Landau-Lifshitz equations for the evolution of spin components read

\begin{equation}
\dot{x}=\frac{k_x}{m}-2\lambda S_y 
\label{eq1}
\end{equation}
\begin{equation}
\dot{y}=\frac{k_y}{m}+2\lambda S_x 
\label{eq2}
\end{equation}
\begin{equation}
\dot{k}_x(t)=eE_x
\label{eq3}
\end{equation}
\begin{equation}
\dot{k}_y(t)=0
\label{eq4}
\end{equation}
\begin{equation}
\dot{{\bf S}}=-{\bf \Delta} \times {\bf S} 
\label{eq5}
\end{equation}
where ${\bf \Delta}$ is the effective magnetic field acting on the spin. 
 From the Hamiltonian (\ref{hcl1}) we can read this effective magnetic
field acting on the particle with the momentum ${\bf k}=(k_{x},k_{y})$
\begin{equation}
{\bf \Delta }=-2(\lambda k_{y},-\lambda k_{x}, h)
  \label{d1}
\end{equation}
The absolute magnitude of this field is
\begin{equation}
|{\bf \Delta }|=\sqrt{4(k^{2}\lambda ^{2}+h^2)}
  \label{d12}
\end{equation}

 Assuming that the electric field $E_x$ is weak, the variation of the momentum and hence of the effective magnetic field is very slow and,
  if at the initial moment the classical spin is directed along the magnetic field, it will  follow the direction of this slowly varying field.
This adiabatic approximation follows from the Landau-Lifshitz equations and yields an approximate
solution for (\ref{eq5}) given by ${\bf S}^{(a)}(t)=\pm |S|\frac{{\bf \Delta}(t)}{|{\bf \Delta}|}$, where 
"$+$" and "$-$" correspond to the initial direction along or opposite to the magnetic field respectively. 
Substituting this time dependence of the spin direction into equations for particle velocities one can find that in the adiabatic approximation
\begin{equation}
\begin{array}{l}
\dot{x}=\frac{\partial \epsilon_0({\bf k})}{\partial k_x}\\
\\
\dot{y}=\frac{\partial \epsilon_0({\bf k})}{\partial k_y}
\end{array}
\label{adxy}
\end{equation}
where $ \epsilon_0({\bf k})=\frac{k^2}{2m} \mp \sqrt{(k\lambda)^2+h^2}$, i.e. a particle with spin along the field $\Delta$ has energy dispersion
as an  electron in the major band of R2DEG.
In the strict adiabatic approximation of this classical Hamiltonian particles do not exhibit 
the anomalous velocity seen in the second term of the group velocity of the quantum-mechanical wavepacket 
dynamics described by Eq. (2).

However, this classical model does capture the correct contribution to the anomalous Hall effect 
related to the Berry curvature once we go beyond the strict adiabatic approximation and consider instead the
more general linear response. The nonadiabatic corrections describe deflection of the classical 
spin from the direction of the instant magnetic field. They are small but can still be of the first 
order in the electric field and hence affect the transport properties in the linear response. 
Looking for a solution of the Landau-Lifshitz equations (\ref{eq5}) 
in the form ${\bf S}={\bf S}^{(a)}+\delta {\bf S}$ we find that
in the first order in the external electric field the spin direction acquires the component
perpendicular both to the effective field and its derivative \cite{SHE} i.e.

\begin{equation}
\delta {\bf S} = \pm |S|\frac{{\bf \Delta}}{|\bf \Delta|^3} \times \frac{d{\bf \Delta}}{dt}
\label{ahe1}
\end{equation}
If $h \ne 0$ this correction has in-plane components and hence affects the velocity of the particle, i.e.
it gives a contribution to the right hand side of Eq. \ref{eq1} and \ref{eq2}.
Substituting (\ref{d1}), (\ref{d12}) into (\ref{ahe1}) we find
\begin{equation}
\delta S_{x} (({\bf k})=\pm |S|\frac{\lambda h \dot{k}_x}{2[(k\lambda)^2+h^2]^{3/2}}
  \label{nzp1}
\end{equation}
\begin{equation}
\delta S_{y} (({\bf k})=\mp |S|\frac{\lambda h \dot{k}_y}{2[(k\lambda)^2+h^2]^{3/2}}
  \label{nzp2}
\end{equation}
Then substituting (\ref{nzp1}) and (\ref{nzp2}) into (\ref{eq1}), (\ref{eq2}) for $|S|=1/2$
we find equations identical to the wave packet ones with the Berry curvature given in Eqs. (\ref{om1}) and (\ref{bc1}) 

We emphasize that although being an interesting demonstration of the effect,
the model of classical charged particles with classical spins cannot always be 
employed in calculations of the Berry curvature and quantum
corrections  for other types of the Hamiltonian may appear. \cite{Murak}

\end{document}